\documentstyle[prl,aps,multicol]{revtex}
\begin{document}
\twocolumn[ 
\hsize\textwidth\columnwidth\hsize\csname@twocolumnfalse\endcsname 

\title{Structural characterization of 
YBa$_{2}$Cu$_{3}$O$_{7-\delta}$/Y$_{2}$O$_{3}$ composite films}
\author{P.R. Broussard}
\address{Naval Research Lab, Washington, DC 20375}
\author{M.A. Wall}
\address{Lawrence Livermore National Laboratory, Livermore, CA 94550}
\author{J. Talvacchio}
\address{Northrop Grumman Science and Technology Center, Pittsburgh, PA  15235}
\bigskip 
\maketitle
\begin{abstract}
Using 4-circle x-ray diffraction and transmission electron microscopy 
we have studied the microstructure 
and in-plane orientation of the phases present in thin film composite 
mixtures of YBa$_{2}$¥Cu$_{3}$¥O$_{7-\delta}$ and Y$_{2}$¥O$_{3}$. 
We see a high degree of in-plane orientation 
and have verified a previous prediction for the in-plane order of 
Y$_{2}$¥BaCuO$_{5}$¥ on (110) MgO.  Transmission electron microscopy 
shows the composite films to be a mixture of two phases, with YBCO 
grain sizes of $\approx$ 1 $\mu$m.  We have also compared our observations of 
the in-plane order to the predictions of a modified near coincidence 
site lattice model.
\end{abstract}

\pacs{61.16.Bg, 61.10.-i, 74.76.Bz, 68.55.-a}
]

\narrowtext

\section{Introduction}¥
In thin film growth of YBa$_2$Cu$_3$O$_{7-\delta}$ (YBCO), it is useful to 
understand the nature of the epitaxial relationships present between 
impurity phases in the films.  This knowledge provides a guide for 
finding these phases by x-ray diffraction, as seen in the study of 
CuO in YBCO films by Watson et al.\cite{Watson}¥   Also, in a previous 
report on thin film composites of YBCO and yttria,\cite{Broussard1}¥  
we predicted the in-plane orientation of the phase Y$_{2}$¥BaCuO$_{5}$ (Y-211) 
which was found in composites grown on (110) MgO.  In this work, 
we have studied the microstructure and in-plane orientation of the 
phases present in thin film composites of YBCO and yttria grown on 
(100) and (110) MgO.  
In particular, we have studied how the YBCO, yttria, and Y-211 phases 
grow in relation to the underlying substrate, to understand how 
these phases appear as impurities in YBCO and to verify our previous prediction.

\section{Sample Preparation}
The samples were grown by off-axis sputtering onto commercially polished 
(100) and (110) MgO substrates.\cite{Broussard1,Broussard2} YBCO and yttria 
were cosputtered in a 100 mTorr gas mixture of 80\%  argon and 20\%  oxygen at 
a total growth rate of $\approx$ 440 {\AA}/hour.  Under the growth conditions used they would 
ideally be composed of 91\%  YBCO and 9\% yttria by volume.  The substrates were at a 
temperature of $\approx$ 700 $^{\circ}¥$C, and after the growth were furnace 
cooled in 100 
Torr of oxygen.  Film thicknesses were $\approx$ 1900 {\AA}.  The samples were studied 
by 4-circle x-ray diffraction (XRD) and transmission electron microscopy 
(TEM).  The XRD was carried out on a Phillips 4-circle diffractometer.
The orientation of the films was 
determined by standard $\theta-2\theta$ scans with diffraction along 
the sample normal.  The crystal quality was studied using rocking curves 
for diffraction along and at various angles to the sample normal.  In-plane 
order was measured by taking $\phi$ scans where the diffraction vector 
for a particular value of $2\theta$ is at an angle (90-$\chi$) to the sample 
normal and $\phi$ is the angle of rotation about the sample normal.  
The TEM analysis was performed on a JEOL 200CX 
microscope using conventional bright field (BF) and selected area 
diffraction (SAD) conditions on plan view specimens.

To interpret the in-plane order observed and compare lattice matching 
of the phases to the substrate and to each other, previous work has 
used a near coincidence site lattice theory (NCSL).\cite{Hwang}  In this 
model (for the case when the lattice direction along the growth direction is already 
determined)
the mismatch between the film surface mesh and the substrate surface 
mesh is given as the percentage difference of the respective position 
vectors with respect to their average lengths.  Since this approach only 
compares discrete orientations between the two meshes, we have used a 
computer model to study the matching of an N x N mesh of 
an overlayer onto the substrate mesh for arbitrary angles between the two 
meshes.  We have a rectangular overlayer with lattice constants a and b 
in the directions $\hat{x}¥$¥ and $\hat{y}$¥, and a rectangular substrate 
mesh with corresponding lattice constants c and d in the directions 
$\hat{x'}$ and $\hat{y'}$, 
where the primed system is rotated by an angle $\theta$ with respect to 
the unprimed system.  Then for each overlayer lattice point,
${\vec{r}}_{i,j}^{o} = ia\hat{x} + jb\hat{y}$, the program 
finds the corresponding closest lattice point in the substrate layer,
${\vec{r}}_{i,j}^{s} = {k}_{i}c\hat{x'} + {l}_{j} d\hat{y'}$, 
computes the strain for that point, and averages over all overlayer points 
(excluding the origin):
\begin{equation}
Strain = {\frac{1}{{N}^{2} - 1}}\sum\limits_{i,j = 0 \atop 
(i,j)\ \ne (0,0)}^{N} {\frac{2\left|
{{\vec{r}}_{i,j}^{o} - {\vec{r}}_{i,j}^{s}}\right|}{\left|
{{\vec{r}}_{i,j}^{o}}\right| + \left|{{\vec{r}}_{i,j}^{s}}\right|}}.
\label{min}
\end{equation}
We then look for minima in the strain as a function of angle between the 
lattices that will indicate a preferred orientation of the overlayer 
on the substrate.

\section{Results}¥

We 
first consider the composite on (100) MgO, which from $\theta-2\theta$ XRD shows the 
presence of YBCO and yttria in the film.  The YBCO phase is c-axis 
oriented with a lattice constant of 11.714 $\pm$ 0.004 {\AA}.  The rocking curve width for 
the YBCO 005 peak is 0.64$^{\circ}¥$, with a comparable width for the YBCO 
309 peak.  The yttria in 
the film is oriented (001) with a lattice constant of 10.594 $\pm$ 
0.004 {\AA}, with a rocking curve width for the yttria 004 
peak of 0.98$^{\circ}¥$.  The width for the yttria 226 peak is 
comparable to that of the 004 peak.  

In Fig. \ref{XRD 100} we present plots of x-ray intensity versus 
$\phi$ for the MgO 402, 
YBCO 309 peak, and the yttria 226 peak.  We find in contrast to the case of 
pure YBCO on (100) MgO\cite{Williams} that the YBCO in the composite film is primarily 
oriented at the 45$^{\circ}¥$ orientation, or [110] YBCO $\parallel$ [010] MgO.  
A smaller 
amount of YBCO is in the usual Òcube-on-cubeÓ orientation.  We find from 
the $\phi$ 
scans that the bulk of the yttria is oriented as Òcube-on-cubeÓ, or [100] 
yttria $\parallel$ [010] MgO, with a small amount oriented at 45$^{\circ}¥$, or [110] yttria 
$\parallel$ [010] MgO.  We also see that the mosaic spread in the 
$\phi$ scan is 
substantially broader for the yttria peaks, with the FWHM being $\approx$ 7$^{\circ}¥$, 
but only 
$\approx$ 2.5$^{\circ}¥$ for the YBCO peaks.  This larger peak width is consistent with the 
larger rocking curve width seen for the yttria.  Comparing the YBCO 
$\phi$ scan 
to that for the yttria, we see that the YBCO is oriented with respect to 
the yttria as [110] YBCO $\parallel$ [100] yttria, which was seen in studies of 
yttria on YBCO layers.\cite{Hirata,Ying} We also see that the intensities of the two 
phases track as a function of the angle $\phi$.  This tracking implies that the 
matching for the YBCO present in this sample is predominantly with the 
yttria, not the MgO.

Figure \ref{TEM 100} shows a TEM BF image 
of the composite sample on (100) MgO.  There are two phases clearly 
present: equiaxed, highly-faceted islands approximately 1 $\mu$m in diameter 
partially surrounded by a thin phase.  We see a high degree of connectivity 
between the islands in this sample.  Analysis of SAD images (not shown) 
reveals that the islands are c-axis oriented YBCO and the phase between the 
YBCO grains is (001) oriented yttria.  In addition, SAD analysis shows an in-plane orientation relationship 
of the [101] direction of yttria parallel to the [100] or [010] axis of 
YBCO, which agrees with what was seen in the XRD analysis. The 
variation in contrast among the YBCO islands is due to their varying 
thickness. We  
estimate from the image that the volume fraction of YBCO is 
93 $\pm$ 4\%, with the uncertainty being due to contrast variation.   

Using our lattice model for the case of the composite 
film on (100) MgO, we compared the matching of yttria (001) to MgO (100), 
and because the YBCO in this sample seemed to be oriented with the yttria, 
we also compared the matching of yttria (001) to YBCO (001).  For yttria we 
used a=b=10.6 {\AA}.  For yttria (001) on (100) MgO, we find that there is a 
strong minima at 0$^{\circ}¥$ and another at 23$^{\circ}¥$, but not at 
45$^{\circ}¥$.  (The corresponding NCSL parameters for these lattice 
matchings would be $\sigma_{yttria}$/$\sigma_{MgO}$¥=4/25, 4/29, and 
9/50, respectively, with NCSL lattice mismatches of 0.88\%, 6.6\%, and 6.4\%, 
respectively.)  Experimentally, we 
do not see any yttria misaligned by 23$^{\circ}¥$, but we do observe it at 
45$^{\circ}¥$.  For 
YBCO on yttria, we find a clear minima for the 45$^{\circ}¥$ alignment, 
as seen in 
the films, as well as weaker minima at 0$^{\circ}¥$ (Òcube-on-cubeÓ), 
14$^{\circ}¥$, and 21$^{\circ}¥$, 
which we do not observe.  (The corresponding NCSL parameters for these lattice 
matchings would be $\sigma_{YBCO}$/$\sigma_{yttria}$¥=8/1, 9/1, 41/5, and 
29/4, respectively, with NCSL lattice mismatches of 1.4\%, 7.3\%, 2.6\%, and 
3.5\%, 
respectively.)  From the data and the results of the model, we 
believe the $\phi$ scan in Fig. \ref{XRD 100} can be explained by two types of 
microstructure in this sample.  The first is where yttria is the initial 
phase growing on the MgO, and orients as Òcube-on-cubeÓ, and the YBCO that 
grows subsequently is misoriented by 45$^{\circ}¥$.  This type of 
microstructure is 
the predominant one in the sample.  The second is where YBCO nucleates 
first on the MgO, growing as Òcube-on-cubeÓ, and the yttria that grows 
subsequently is misoriented by 45$^{\circ}¥$.  Careful study of the 
SAD images did 
show evidence of extra diffraction spots, which are most likely due to 
double diffraction.  This would be consistent with the yttria and YBCO 
phases being on top or underneath each other.  

Next, we turn to the 
composite sample on (110) MgO.  As was discussed in Ref. 
\cite{Broussard1}, this sample 
was deposited in the same run as the composite sample on (100) MgO.  
However, XRD shows no yttria in the film, but instead indicates a-axis 
oriented Y-211 material.  The sample was insulating, even though from XRD 
the YBCO present in the film is similar to that for pure YBCO on (110) 
MgO.  We find c-axis oriented YBCO with a c-axis constant of 
11.728 $\pm$ 0.004 {\AA}.  
The rocking curve width for the YBCO 005 peak is 0.6$^{\circ}¥$ (with a comparable 
width for the YBCO 309 peak).  The Y-211 
material in the film is oriented (100) with a lattice parameter of 
7.145 $\pm$ 0.004 {\AA} and has a rocking curve width for 
the Y-211 200 reflection of 0.39$^{\circ}¥$ and 0.46$^{\circ}¥$ for 
the 505 off-axis peak.  
This indicates that the Y-211 material is more ordered than the YBCO in the 
film.  

If we look at the $\phi$ scans for peaks from MgO 240, YBCO 309 and 
Y-211 320, as in Fig. \ref{XRD 110}, we see that the YBCO is oriented 
with [100] YBCO $\parallel$ [001] MgO. The 
Y-211 material is predominantly oriented as [010] Y-211 $\parallel$ [001] MgO, 
with 
about 10\% (by volume) of the Y-211 material oriented as [010] Y-211 $\parallel$ 
[$\bar{1}$10] MgO, 
or a 90$^{\circ}¥$ 
misorientation.  In Ref. \cite {Broussard1}, we predicted that the in-plane 
order of the (100) 
Y-211 material on the (110) MgO substrate would be [010] Y-211 
$\parallel$ [001] MgO 
and [001] Y-211 $\parallel$ [1$\bar{1}$0] MgO.  This is what is seen for the bulk of the Y-211.  

Figure \ref{TEM 110} is a TEM BF image of the composite sample on (110) MgO, which 
clearly shows a two phase structure consisting of irregular-shaped, 
slightly faceted islands of average width 0.5 $\mu$m surrounded by a 
sea/matrix.  Analysis of the SAD images shows the island phase to be c-axis 
oriented YBCO.  Rotational spot splitting is also observed, which results 
from twinning within the islands of YBCO.  The matrix phase is single phase 
a-axis oriented Y-211.  SAD patterns from different parts of the sample had 
the same orientation, indicating the Y-211 grains are single domain 
over the region examined ($\approx$ 2 $\mu$m).  
Figure \ref{TEM 110} shows a 
small amount of connectivity between the YBCO islands, but for the most 
part they are isolated.  This result is consistent with the sample being 
insulating.  

For the composite sample on (110) MgO, we used Eq. 
\ref{min} to model 
the matching of Y-211 (100) to MgO (110) using for Y-211 lattice parameters 
b=12.16 $\pm$ 0.04 {\AA} and c= 5.656 $\pm$ 0.002 {\AA} (measured for this sample).  
We find the strongest 
minima for the predominant orientation of [010] Y-211 $\parallel$ [001] MgO, 
but the 
90$^{\circ}¥$ orientation [010] Y-211 $\parallel$ [$\bar{1}$10] MgO also has a strong minima.  There is also 
a weak minima for 45$^{\circ}¥$ oriented material.  (The corresponding NCSL parameters for these lattice 
matchings would be $(k,l)_{Y-211}$/$(k,l)_{MgO}$¥=(1,0)/(3,0), 
(0,3)/(4,0), and 
(1,1)/(1,2), respectively, with NCSL lattice mismatches of 1.4\%, 0.92\%, and 
6.1\%, 
respectively.)  For the case of Y-211 (100) to 
YBCO (001), we find the best match at a 45$^{\circ}¥$ misorientation, and weaker 
minima for either Y-211 [010] or [001] $\parallel$ YBCO [100].  (The corresponding NCSL parameters for these lattice 
matchings would be $(k,l)_{Y-211}$/$(k,l)_{YBCO}$¥=(0,1)/(1,1), and 
(0,2)/(0,3), respectively, with NCSL lattice mismatches of 5.1\%, and 
0.77\%, 
respectively.)  Since we did not 
observe the 45$^{\circ}¥$ misorientation in the x-ray scans, we conclude that the 
Y-211 and YBCO phases grow on the (110) MgO substrate, unlike the case for 
a composite sample on (100) MgO with YBCO and yttria, where the evidence is 
that the bulk of the YBCO grains grow on yttria, not on the (100) MgO.

\section{Discussion}

One of the major motivations for this work was to understand why different 
phases form on (100) MgO than (110) MgO substrates for composite samples.  
The in-plane ordering we observe suggests that phase formation is driven by 
lattice matching to the substrate.  However, if we use Eq.  1 to predict 
how yttria or Y-211 will grow on the two orientations of MgO, the results 
do not agree with our XRD studies.  For (001) yttria, the model predicts a 
lower strain for growth on (100) MgO compared to (110) MgO.  For a-axis 
oriented Y-211, the model also predicts a lower strain for growth on (100) 
MgO, not (110) MgO, with an in-plane orientation at 45$^{\circ}¥$.  These results 
suggest that if substrate lattice matching is driving the formation of 
Y-211 on (110) MgO, then Y-211 should also form on (100) MgO.  Our studies 
show otherwise. 

Another approach to understanding the growth of Y-211 on 
(110) MgO is to assume it is controlled by the bulk phase diagram for the 
Y-Ba-Cu-O system.  We would estimate from the location of the composite 
sample on (110) MgO in the Y-Ba-Cu-O phase diagram (in Ref. 
\cite{Broussard1}) that there 
should be 49\%  YBCO, 39\%  Y-211 and 12\%  CuO by volume.  However, 
from Fig.  
\ref{TEM 110} 
we find that the volume fraction of YBCO is only 35\% , and neither XRD or 
TEM shows the presence of CuO in the films.  Clearly the growth mechanism 
on (110) MgO is not fully understood.

\section{Conclusions}

We have seen by both XRD and TEM that composite films of YBCO and yttria 
exhibit a great deal of in-plane orientation.  We have successfully 
verified the prediction in Ref. \cite{Broussard1} that composites grown on (110) MgO have 
a-axis oriented Y-211 present with a predominant in-plane orientation of 
[010] Y-211 $\parallel$ [001] MgO, but some fraction of the 
Y-211 phase is oriented at 90$^{\circ}¥$ to this.  The composites on (001) MgO show 
in-plane orientations that indicate two types of microstructure, either 
yttria on YBCO or YBCO on yttria.  The sizes of the YBCO grains are 
$\approx$ 1 $\mu$m 
for the composites on (100) MgO, while on (110) MgO YBCO forms oblong 
islands of up to several microns in length.  Our attempts at studying the 
lattice matching using a modified near coincidence site lattice model did 
show that we could understand the observed in-plane orientations, but could 
not explain the occurrence of different phases growing on (100) and (110) 
MgO.

\begin{figure}
\caption{X-ray intensity (log scale) versus $\phi$ for diffraction 
along the yttria 226, YBa$_{2}$¥Cu$_{3}$¥O$_{7-\delta}$ 309 and the MgO 
402 peaks for the composite sample grown on (100) MgO.  The curves are 
offset for clarity.}
\label{XRD 100}
\end{figure}

\begin{figure}
\caption{Bright field plan view TEM images for the composite sample grown on (100) 
MgO showing YBCO grain structure with yttria at the grain boundaries.  The 
darker faceted islands are the YBCO phase, with the yttria phase showing up 
as white in the grain boundaries.}
\label{TEM 100}
\end{figure}

\begin{figure}
\caption{X-ray intensity (log scale) versus $\phi$ for diffraction along the 
Y$_{2}$Ba¥Cu¥O$_{5}$ 320, YBa$_{2}$¥Cu$_{3}$¥O$_{7-\delta}$ 309 and the MgO 240 peaks for 
the composite sample 
grown on (110) MgO.  The curves are offset for clarity.}
\label{XRD 110}
\end{figure}

\begin{figure}
\caption{Bright field plan view TEM image for the composite sample grown on (110) MgO 
showing the morphology of the sample at low magnification.  The lighter 
region is the Y-211 phase, with the YBCO phase in oblong islands.}
\label{TEM 110}
\end{figure}

\end{document}